\pdfoutput=1
\documentclass[runningheads]{llncs}
\usepackage{graphicx}
\begin{document}
\title{FuzzTheREST: An Intelligent Automated Black-box RESTful API Fuzzer}
\titlerunning{Intelligent Automated Black-box RESTful API Fuzzer}
\author{Tiago Dias\orcidID{0000-0002-1693-7872} \and
Eva Maia\orcidID{0000-0002-8075-531X} \and
Isabel Praça\orcidID{0000-0002-2519-9859}}
\authorrunning{T. Dias et al.}
\institute{Research Group on Intelligent Engineering and Computing for Advanced Innovation and Development (GECAD), Porto School of Engineering (ISEP), 4200-072 Porto, Portugal \\
\email{\{tiada,egm,icp\}@isep.ipp.pt}}
\maketitle              % typeset the header of the contribution
\begin{abstract}
Software's pervasive impact and increasing reliance in the era of digital transformation raise concerns about vulnerabilities, emphasizing the need for software security. Fuzzy testing is a dynamic analysis software testing technique that consists of feeding faulty input data to a System Under Test (SUT) and observing its behavior. Specifically regarding black-box RESTful API testing, recent literature has attempted to automate this technique using heuristics to perform the input search and using the HTTP response status codes for classification. However, most approaches do not keep track of code coverage, which is important to validate the solution. This work introduces a black-box RESTful API fuzzy testing tool that employs Reinforcement Learning (RL) for vulnerability detection. The fuzzer operates via the OpenAPI Specification (OAS) file and a scenarios file, which includes information to communicate with the SUT and the sequences of functionalities to test, respectively. To evaluate its effectiveness, the tool was tested on the Petstore API. The tool found a total of six unique vulnerabilities and achieved 55\% code coverage.

\keywords{Fuzzy Testing \and RESTful APIs \and Vulnerability Detection \and Code Security  \and Artificial Intelligence }
\end{abstract}

\section{Introduction}
\label{sec:intro}

Software has had a positive impact over the years and its adherence continues to grow. The ubiquity of information technology, coupled with the seamless availability of web services, such as REST, has made software much more accessible in today's interconnected landscape. However, considering its prevalence, the existence of software vulnerabilities is worrying. Nowadays, people depend on technology, and as such, poor-quality software can have detrimental effects. Considering that software is the backbone of many modern operations, ensuring software reliability and security is crucial~\cite{Malhotra2020,Shen2020,Ghaffarian2017}. 

Software testing is one of the main contributors to software quality capable of answering the validation and verification standard~\cite{Maltaie2020}. Software testing encompasses a very extensive taxonomy of methods, each focusing on improving different software quality attributes~\cite{Ganney2020}. One such method is vulnerability detection, which consists of spotting and addressing weaknesses in software code. Vulnerabilities are typically spotted following a certain type of analysis: (i) static and/or (ii) dynamic. The first one consists of analyzing the source code itself, whilst the second consists of analyzing the behavior of the SUT by feeding it varied input data~\cite{Malhotra2020,Shen2020,Ghaffarian2017,Shafiq2021,Zhu2022}. Fuzzy testing is a dynamic analysis vulnerability detection technique that consists of testing a SUT with faulty input and monitoring its behavior. In recent research, multiple works have attempted to reduce the search space of the inputs for efficient fuzzing, with most approaches relying on heuristics~\cite{constrainedData}. However, most works do not measure code coverage in this setting.

This paper introduces a black-box RESTful API fuzzy testing tool that leverages RL to perform efficient fuzzy testing. The fuzzer interacts with the SUT in an interoperable manner by using the OAS file and allows the configuration of scenarios, which are functionality execution sequences. The role of the RL in this context is to guide the input generation, by learning the context of the SUT relying on its HTTP responses. A vulnerability report is provided along with the RL agent's performance on the SUT making the intelligent system highly explainable as well.

This paper is organized as follows. Section~\ref{sec:SoTA} provides an analysis of related works. Section~\ref{sec:proposal} details the fuzzer's implementation. Section~\ref{sec:case_study} presents the experimental setup, demonstrates the suitability of the tool in the case study API, and discusses the results. Finally, Section~\ref{sec:conclusion} summarises and outlines the key takeaways of the work, describing the current ongoing improvements to the fuzzer.
\section{State-of-the-Art}
\label{sec:SoTA}
Fuzzy testing is a vulnerability detection technique that has shown remarkable results in identifying software defects, particularly when automated~\cite{Shen2021,Bhme2022,Hazimeh2020,Fioraldi2022}. This technique involves generating faulty input to feed the functionalities of a SUT and monitoring its behavior according to the actual requirements and performance of the system~\cite{Zhu2022,Shen2020}. Since it consists of trying different inputs to find vulnerabilities, this method is prone to low code coverage due to randomness, and inefficient execution due to redundant tests~\cite{Shen2021} leading to a higher rate of missed vulnerabilities. However, it is generally more accurate and has fewer false alarms than static analysis~\cite{Zhu2022}.

Nowadays, web APIs are the de-facto standard for Software integration~\cite{Martin-Lopez2019}, being most software made available via RESTful APIs. As RESTful APIs gain momentum, so does their testing. Recent advances in the literature show that black-box testing of RESTful APIs has outputted effective results and is capable of contributing to the validity, reliability, and correctness of these systems.

Atlidakis et al.~\cite{atlidakis2019restler} presented RESTler, a REST API fuzzy testing tool. The tool analyses the API specifications and generates a sequence of requests by inferring order dependencies and analyzing dynamic feedback of prior tests. Their tool was capable of detecting 28 bugs in various cloud services, which were confirmed and tested by service owners. Ed-douibi et al.~\cite{EdDouibi2018AutomaticGO} also presented an automated REST API testing tool that relied on OpenAPI Specifications (OAS) to generate specification-based test cases to ensure APIs meet the requirements. Their tool was validated on 91 APIs, and the results showed that the generated test cases cover, on average, 76.5\% of the elements included in the definitions, with 40\% of the tested APIs failing. Karlsson et al.~\cite{Karlsson2019} attempted automatic property-based tests generated from OAS documentation, where the input was randomly generated. Haraldsson et al.~\cite{Haraldsson2017} collected public data and made it faulty to produce valuable test cases. 

More recently, research has attempted to leverage RL to solve the search-based problem of finding the right input and combination of use cases. Sameer et al.~\cite{Sameer2020} utilized RL to enhance valid test input generation, addressing the challenge of generating valid inputs for parameterized test drivers like web APIs. They introduce a tabular guide based on the Monte Carlo Control (MCC) learner, which constrains the input space within a random generator. While their system may face scalability issues due to the complexity of the MCC algorithm, their tool, demonstrated improved average time and reliability in bug detection in comparison to other tools, though it was unable to detect all bugs. Similarly, Zheng et al.~\cite{Zheng2021} introduced WebExplor, an automated web testing framework employing curiosity-driven reinforcement learning to generate action sequences for end-to-end web testing. The framework adopts a Use Case testing approach, recognizing that a sequence of system functionalities can reveal unexpected defects. To address local optima, they incorporate a Deterministic Finite Automaton to track RL exploration. WebExplor successfully identified 3466 exceptions and errors in a real-world commercial web application. 

Although these approaches can diminish the input space, and produce realistic faulty test data leading to the discovery of vulnerabilities, most do not provide code coverage metrics to validate the solutions and only describe the errors or vulnerabilities found in numbers, without explaining them. This work tackles these issues by presenting a novel fuzzer that uses a custom RL-based approach to guide input generation.
\section{FuzzTheREST: A TestLab Module}
\label{sec:proposal}

This work presents FuzzTheREST, an automated black-box RESTful API fuzzer that leverages RL to strategically generate input values. The tool integrates the TestLab ecosystem, which is an automated software testing framework that integrates a diverse range of testing approaches across various levels and viewpoints~\cite{Dias2023}. FuzzTheREST's approach utilizes as main data source the OAS file for interoperability purposes. Several mutation methods were implemented to generate faulty input for the fuzzy tests. When successful, these result in the SUT's anomalous response, which the RL algorithm learns to improve input generation. In addition, the fuzzer documents the findings and performance of the RL, granting visibility into the process. This fuzzer sets itself apart through its novel RL algorithmic approach, explainability, adaptability, and interoperability. By eliminating the need for in-depth source code knowledge, it streamlines the testing process, making it more accessible.

\subsection{Reinforcement Learning Design}

\begin{figure}[h]
\centering
\includegraphics[width=0.65\textwidth]{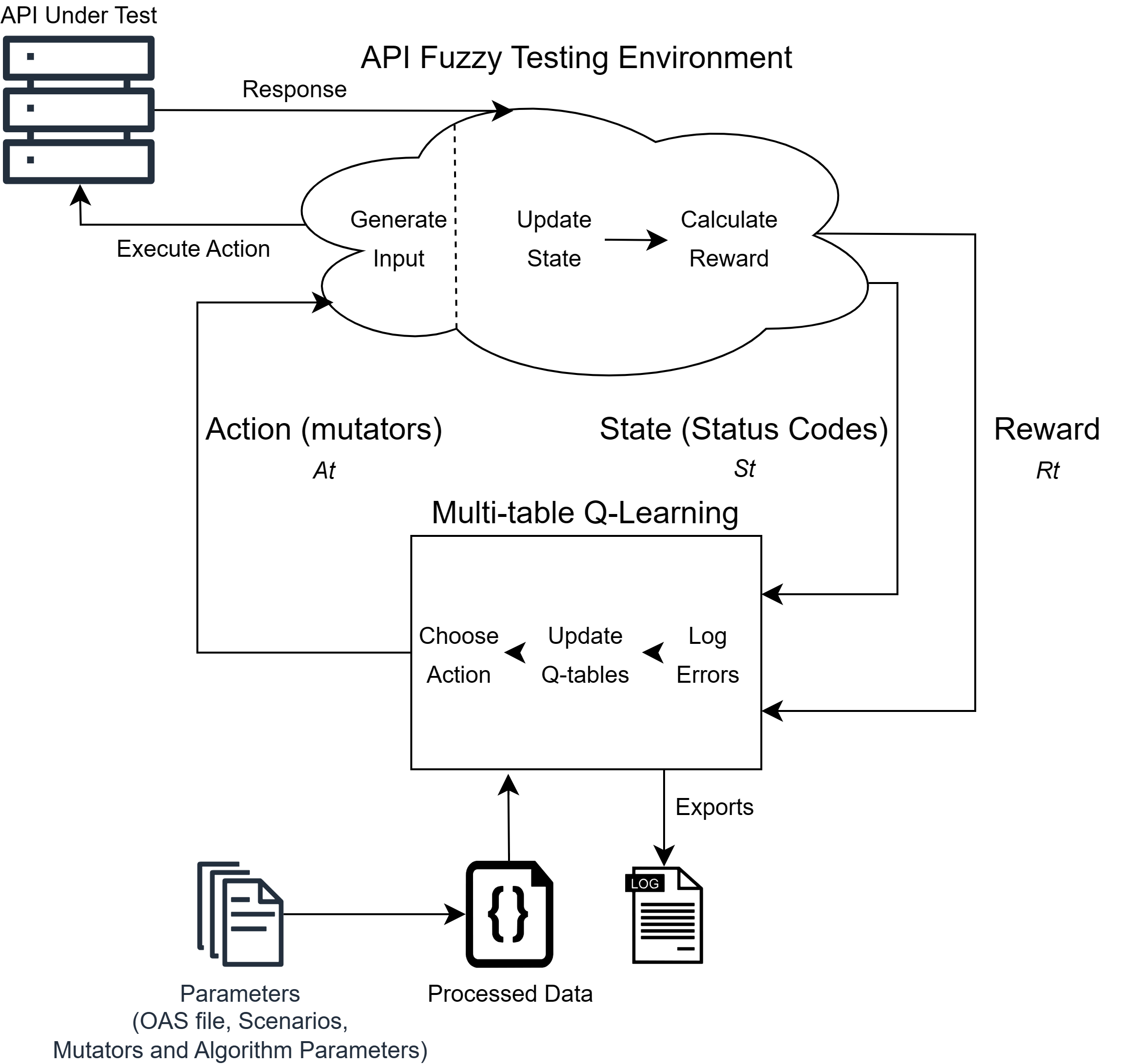}
\caption{Reinforcement Learning Environment and Agent Overview}
\label{fig:rl_overview}
\end{figure}

This section describes meticulously the design and execution of the novel RL algorithmic approach employed in the fuzzer for improved input generation. Fig~\ref{fig:rl_overview} depicts the custom environment and the agent characteristics, interactions, and architecture. The following subsections describe the data acquisition process, RL environment, and Agent's design, showcased in Fig~\ref{fig:rl_overview}.

\subsubsection{Data Sources and Processing.}

The fuzzer requires four main sources of information to operate: (i) the OAS file, containing information to communicate with the SUT, (ii) the scenarios file, containing functionality execution sequences, (iii) the mutators, responsible for generating the faulty input, and (iv) the algorithm parameters, which will impact the learning process of the agent, therefore affecting its performance and the results obtained during testing.

The OAS file is divided into three main sections: (i) API information, (ii) API paths and operations, and (iii) reusable components for data schemas and security schemes. Since the OAS file contains a huge corpus of unnecessary data, before training the RL algorithm, this file was processed for dimensional reduction.

\begin{figure}[h]
\centering
\includegraphics[width=0.70\textwidth]{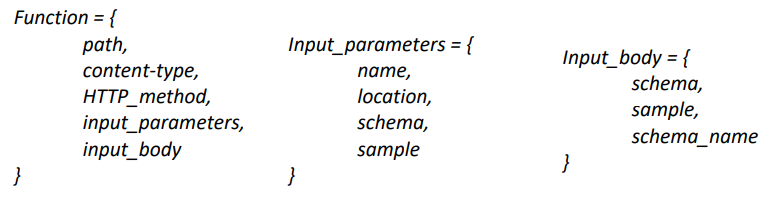}
\caption{Proposed Fuzzer's Data Schema}
\label{fig:data_schema}
\end{figure}

The extracted data is the one shown in Fig.~\ref{fig:data_schema}, following a JSON notation. The~\texttt{Function} contains the request type, header, and payload information of each API function. The~\texttt{Input\_parameters} store each function's query parameters and the~\texttt{Input\_body} the request body schema. The fuzzer is implemented to deal currently with five different datatypes: (i) integer, (ii) float, (iii) boolean, (iv) string, (v) and byte. The~\texttt{sample} field in both~\texttt{Input\_parameters} and~\texttt{Input\_body} represents an instantiated schema. This value can be randomly generated or mutated via the mutation methods. Eight mutation methods were implemented inspired in successful fuzzers like AFL and libFuzzer: (i) bit flips, (ii) byte shuffling, (iii) byte injection/deletion, (iv) bytes substitution, (v) truncation, (vi) dictionary, (vii) arithmetic operations, and (viii) random generation. 

\subsubsection{Reinforcement Learning Environment and Agent}

The RL environment is composed of the observation and action space. The states of the observation space are the HTTP status codes. The action space is defined by the actions that the agent can take. In this context, it corresponds to the mutation methods previously mentioned. Considering the nature of the mutation methods, there is a relationship of one-to-many between the datatypes and actions.

The RL agent interacts with the environment, by acting according to the current state. In this environment, the action is making an HTTP request with the faulty input to the SUT, which generates a response containing the HTTP status code. The status is used to assign a reward, which can be positive or negative depending on the outcome, to the agent. In this work, rewards were modeled around HTTP status codes: 1XX rewards 0 points, 2XX and 3XX rewards 5 points, 4XX takes 20 points, and 5XX, grants 10 points. This action and reward system allows the agent to learn over time. Nonetheless, it is important to choose a suitable agent to interact with such a specific environment. 

An agent is an algorithm that learns from the environment and makes decisions to maximize cumulative rewards. This is done by figuring out the best possible actions to take given its current state. The agent is a Multi-table Q-Learning algorithm~\cite{kantasewi2019} that is well-suited to the environment developed. The Multi-table approach was motivated by the relationship between datatypes and actions, assigning each datatype its own Q-Table. To optimize learning, an epsilon-greedy policy with exploration decay was implemented to balance exploitation and exploration.

Fig.~\ref{fig:rl_overview} depicts the interactions between the agent and the environment during the training process. For a parameterized amount of episodes, the agent engages with the environment, recording outcomes and updating Q-values in the Q-tables. These values measure the success of each action in a given state and are calculated using the Bellman Equation. The agent then applies the policy to decide the next action. In each training episode, the agent can perform a parameterized amount of steps in the environment. However, it may end sooner if a 5XX HTTP status code is encountered. In this work, each agent only tests one single function. As such, if the function being tested consists of a creational one (e.g.: create user), then the agent uses substring identification and word inflection to store the unique identifiers that are generated so that others may use that information, ensuring the dependency between different functions.
\section{Petstore API Case Study}
\label{sec:case_study}

To demonstrate the proposed fuzzer's effectiveness, a case study was conducted on the Petstore API~\cite{Petstore}. The Petstore API is an illustrative Java-based RESTful API, designed around the concept of a pet store business, which includes three main entities: (i) Pets, (ii) Orders, (iii) and Users. For each entity, there are several variants of Create, Read, Update, and Delete (CRUD) methods. For the case study, three distinct scenarios targeting the functions of each entity were crafted in the CRUD order. This API is often used by many literature works in the testing field~\cite{Liu2022,Corradini2021,Mahmood2022,Corradini2022}, and as such it was considered most relevant for the benchmark.

The Petstore API was deployed to a Docker container, to simulate a production environment, and a JaCoCo agent was set up for real-time code coverage. The API was tested using the fuzzer with the parameters described in Table~\ref{tab:param_values}. In bold are the parameters that had the best performance.

\begin{table}[h]
\centering
\caption{Agent's Hyperparameters}\label{tab:param_values}
\begin{tabular}{|p{5cm}|p{7cm}|}
\hline
\textbf{Parameter name} & \textbf{Value} \\ \hline
Number of max. steps & 5 or \textbf{10} \\ \hline
Number of episodes & 50 or 100 or 200 or \textbf{500} \\ \hline
Exploration rate & \textbf{1} or 0.8 \\ \hline
Exploration decay rate & 0.01 \\ \hline
\end{tabular}
\end{table}

\begin{figure}[h]
\centering
\includegraphics[width=\textwidth]{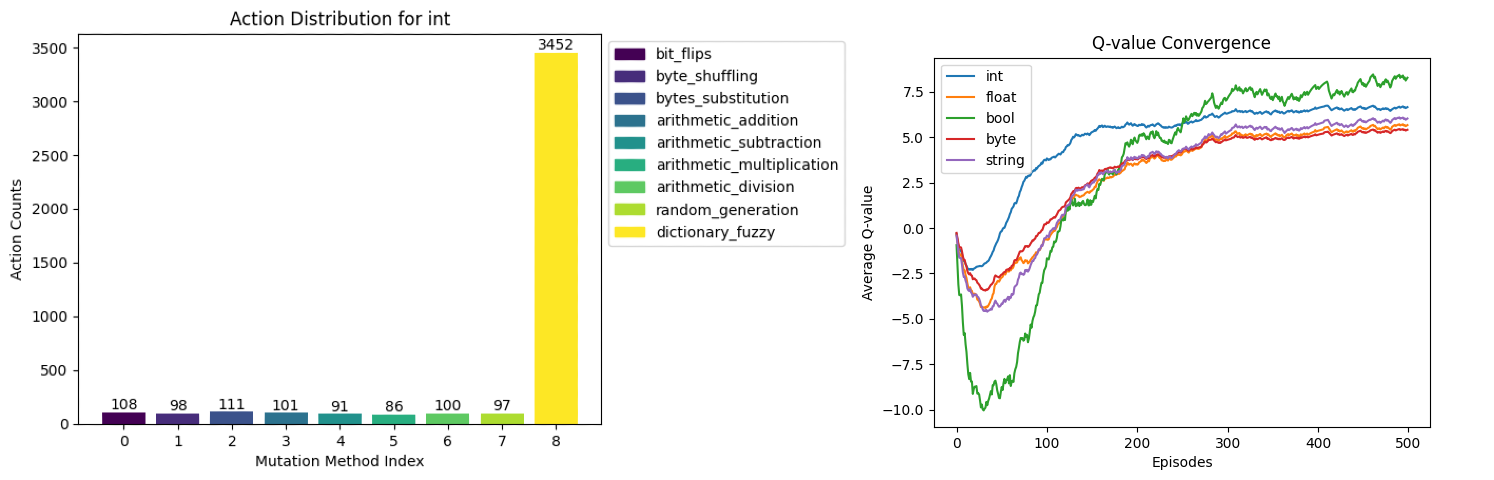}
\caption{GetPetById Agent performance}
\label{fig:get_pet}
\end{figure}

Harnessing the parameters detailed in Table~\ref{tab:param_values}, a suite of 18 agents, one for each tested function, were deployed and trained. The results show that agents testing functions where no vulnerabilities were found quickly reach a plateau. However, this was not true for all functions. For instance, the results achieved in the GetPetById function, depicted in Fig.~\ref{fig:get_pet}, which represent the convergence of Q-Values and the number of received HTTP status codes that range from 1XX to 5XX, show that the agent is exploring the environment and is just barely making any successful requests, with most responses being in the 4XX range. Nonetheless, after a few training iterations, the agent learns which input data works best to interact with the SUT, leading to an increase in success and therefore receiving multiple HTTP responses with status 2XX and 5XX. This pattern replicates in multiple functions, exhibiting the learning process of the agents. A deeper analysis showed that initially, the agent was trying to fetch non-existing pets, and only when it learned that it should choose valid unique identifiers did it start to successfully communicate with the SUT. Since unique identifiers are integers, the action distribution plot corroborates with this observation, reinforcing the assertion that the automated requests are indeed yielding productive outcomes.

The results also show that depending on the complexity of functions the agent takes longer to learn better, as its interaction requires learning more details. An example of this was the findPetsByStatus which requires the pet status attribute to be of a specific range of values.

From an analysis of the vulnerability reports extracted from each function, out of the 2549 requests that resulted in a 5XX status code, six unique vulnerabilities were identified. Table~\ref{tab:vulnerability_status} summarizes the findings.

\begin{table}[h]
\centering
\caption{Petstore Vulnerabilities Found}\label{tab:vulnerability_status}
\begin{tabular}{|p{2.5cm}|p{1.5cm}|p{4.6cm}|p{3.8cm}|}
\hline
\textbf{Vulnerability} & \textbf{Status Code} & \textbf{Description} & \textbf{Faulty Framework} \\ \hline
String to Number & 500 & Conversion error from string input to number. & Not applicable \\ \hline
Malicious Content & 500 & Issue with input validation or security checks. & Spring boot security framework \\ \hline
Illegal Surrogate Pair & 500 & Error due to illegal surrogate pair. & com.fasterxml.jackson.core \\ \hline
Unmatched Surrogate Pair & 500 & Second part of surrogate pair missing. & com.fasterxml.jackson.core \\ \hline
Invalid White Space & 200 & Invalid white space makes fetching certain objects impossible. & com.fasterxml.jackson.core \\ \hline
Fetch Operation Error & 500 & Internal server error due to unexpected status for pet records. & Not applicable \\ \hline
\end{tabular}
\end{table}

The reports highlight critical issues within the SUT. Firstly, there's an erroneous string-to-number conversion in the source code. Additionally, multiple functionalities generate 5XX error codes due to potentially malicious content in URLs. Three vulnerabilities found show issues with illegal character combinations resulting in 5XX and surprisingly one 2XX. This is because malformed inputs were recorded but are unable to be fetched. The report shows this problem stems from a deficiency in the \textit{jackson.core} framework. Similarly, the pets recorded with a status outside of the three possible values were unable to be retrieved. 

The Petstore API vulnerabilities documented in other works~\cite{Mahmood2022} were also identified by the proposed fuzzer. However, unlike the proposed tool, most scientific works that present testing tools, do not describe the errors found and mostly focus on the number of received 5XX HTTP error code~\cite{Liu2022,Corradini2022,Viglianisi2020,Tokos2023}

To validate the effectiveness of the exploration-exploitation inherent to the RL approach implemented, the code coverage was also collected during the testing. 

\begin{figure}[h]
\centering
\includegraphics[width=\textwidth]{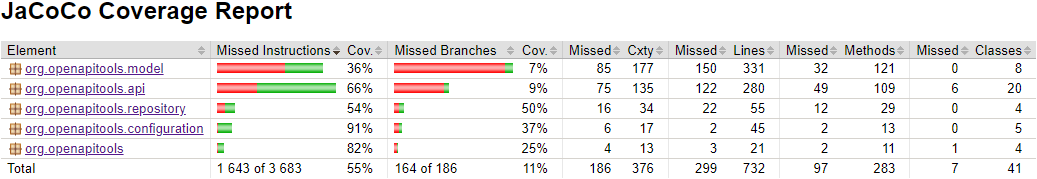}
\caption{JaCoCo Coverage Report}
\label{fig:cov}
\end{figure}

The code coverage report depicted in Fig.~\ref{fig:cov}, shows that 55\% of code was traversed. Naturally, due to the nature of fuzzy testing, some paths may have been missed, which of course reflects on the code coverage achieved. Nonetheless, after further analysis it was noticed that several classes that had defined methods, which are included in the coverage calculation, are never utilized, meaning that such code will never be executed nor tested but will still affect the coverage. As such, if the unused code was removed, the coverage would increase as many methods contain multiple instructions that are never traversed.
\section{Conclusions}
\label{sec:conclusion}

This work presents an RL-based black-box RESTful API fuzzer that tackles input generation, exhaustive testing, and function dependency problems. The system was evaluated in a case study API and was able to uncover six unique vulnerabilities whilst achieving considerable code coverage, highlighting its learning capability. The experiment also shows that the HTTP status codes cannot be trusted alone for identifying vulnerabilities, which has been used in many literature works. Notably, the tool's usage did not impose significant time constraints on the testing process, rendering it a valuable asset for software testing. Further improvements include increasing data types, mutators, and algorithms, and improvements to the unique identifier storing method to make it suitable for all APIs. Incorporating coverage as a decision metric during testing could provide valuable guidance for balancing exploration and exploitation.

\subsubsection{Acknowledgements.} The present work has received funding from \linebreak UIDP/00760/2020 and UIDB/00760/2020.

\bibliographystyle{splncs04}
\bibliography{bibliography}

\begin{thebibliography}{10}
\providecommand{\url}[1]{\texttt{#1}}
\providecommand{\urlprefix}{URL }
\providecommand{\doi}[1]{https://doi.org/#1}

\bibitem{Petstore}
{S}wagger {U}{I} --- petstore.swagger.io. \url{https://petstore.swagger.io/}

\bibitem{atlidakis2019restler}
Atlidakis, V., et~al.: Restler: Stateful rest api fuzzing. In: 2019 IEEE/ACM
  41st International Conference on Software Engineering (ICSE). pp. 748--758
  (2019). \doi{10.1109/ICSE.2019.00083}

\bibitem{Bhme2022}
B\"{o}hme, M., Szekeres, L., Metzman, J.: On the reliability of coverage-based
  fuzzer benchmarking. In: Proceedings of the 44th International Conference on
  Software Engineering. {ACM} (May 2022). \doi{10.1145/3510003.3510230}

\bibitem{Corradini2021}
Corradini, D., et~al.: Restats: A test coverage tool for restful apis.
  Proceedings - 2021 IEEE International Conference on Software Maintenance and
  Evolution, ICSME 2021 pp. 594--598 (2021).
  \doi{10.1109/ICSME52107.2021.00063}

\bibitem{Corradini2022}
Corradini, D., et~al.: Automated black-box testing of nominal and error
  scenarios in restful apis. Software Testing, Verification and Reliability
  \textbf{32},  e1808 (8 2022). \doi{10.1002/STVR.1808}

\bibitem{Dias2023}
Dias, T., et~al.: Testlab: An intelligent automated software testing framework.
  In: Mehmood, R., Alves, V., Pra{\c{c}}a, I., Wikarek, J.,
  Parra-Dom{\'i}nguez, J., Loukanova, R., de~Miguel, I., Pinto, T., Nunes, R.,
  Ricca, M. (eds.) Distributed Computing and Artificial Intelligence, Special
  Sessions I, 20th International Conference. pp. 355--364. Springer Nature
  Switzerland, Cham (2023)

\bibitem{EdDouibi2018AutomaticGO}
Ed-Douibi, H., et~al.: Automatic generation of test cases for rest apis: A
  specification-based approach. 2018 IEEE 22nd International Enterprise
  Distributed Object Computing Conference (EDOC) pp. 181--190 (2018)

\bibitem{Fioraldi2022}
Fioraldi, A., et~al.: Dissecting american fuzzy lop -- a fuzzbench evaluation
  (2022). \doi{10.13140/RG.2.2.13803.82722}

\bibitem{Ganney2020}
Ganney, P.S., et~al.: Software engineering, p. 131–168. Elsevier (2020).
  \doi{10.1016/b978-0-08-102694-6.00009-7}

\bibitem{Ghaffarian2017}
Ghaffarian, S.M., et~al.: Software vulnerability analysis and discovery using
  machine-learning and data-mining techniques: A survey. ACM Computing Surveys
  \textbf{50}(4),  1–36 (Aug 2017). \doi{10.1145/3092566}

\bibitem{Haraldsson2017}
Haraldsson, S.O., et~al.: The use of automatic test data generation for genetic
  improvement in a live system. In: 2017 IEEE/ACM 10th International Workshop
  on Search-Based Software Testing (SBST). pp. 28--31 (2017).
  \doi{10.1109/SBST.2017.10}

\bibitem{Hazimeh2020}
Hazimeh, A., et~al.: Magma. Proceedings of the {ACM} on Measurement and
  Analysis of Computing Systems  \textbf{4}(3),  1--29 (Nov 2020).
  \doi{10.1145/3428334}

\bibitem{kantasewi2019}
Kantasewi, N., et~al.: Multi q-table q-learning. 10th International Conference
  on Information and Communication Technology for Embedded Systems, IC-ICTES
  2019 - Proceedings  (4 2019). \doi{10.1109/ICTEMSYS.2019.8695963}

\bibitem{Karlsson2019}
Karlsson, S., et~al.: Quickrest: Property-based test generation of
  openapi-described restful apis (2019). \doi{10.48550/ARXIV.1912.09686}

\bibitem{Liu2022}
Liu, Y., et~al.: Morest: Model-based restful api testing with execution
  feedback. p. 1406–1417. Association for Computing Machinery (2022).
  \doi{10.1145/3510003.3510133}

\bibitem{Maltaie2020}
M.~Altaie, A., et~al.: Verification and validation of a software: A review of
  the literature. Iraqi Journal for Computers and Informatics  \textbf{46}(1),
  40–47 (Jun 2020). \doi{10.25195/ijci.v46i1.249}

\bibitem{Mahmood2022}
Mahmood, R., et~al.: A framework for automated api fuzzing at enterprise scale.
  Proceedings - 2022 IEEE 15th International Conference on Software Testing,
  Verification and Validation, ICST 2022 pp. 377--388 (2022).
  \doi{10.1109/ICST53961.2022.00018}

\bibitem{Malhotra2020}
Malhotra, R., et~al.: A systematic review on application of deep learning
  techniques for software quality predictive modeling. In: 2020 International
  Conference on Computational Performance Evaluation (ComPE). IEEE (Jul 2020).
  \doi{10.1109/compe49325.2020.9200103}

\bibitem{Martin-Lopez2019}
Martin-Lopez, A., et~al.: Test coverage criteria for restful web apis. p.
  15–21. Association for Computing Machinery (2019).
  \doi{10.1145/3340433.3342822}

\bibitem{Sameer2020}
Reddy, S., et~al.: Quickly generating diverse valid test inputs with
  reinforcement learning. p. 1410–1421. Association for Computing Machinery
  (2020). \doi{10.1145/3377811.3380399}

\bibitem{Shafiq2021}
Shafiq, S., et~al.: A literature review of using machine learning in software
  development life cycle stages. IEEE Access  \textbf{9},  140896–140920
  (2021). \doi{10.1109/access.2021.3119746}

\bibitem{Shen2021}
Shen, Q., et~al.: A systematic review of fuzzy testing for information systems
  and applications. pp. 156--162. IEEE (12 2021).
  \doi{10.1109/CECIT53797.2021.00035}

\bibitem{Shen2020}
Shen, Z., et~al.: A survey of automatic software vulnerability detection,
  program repair, and defect prediction techniques. Security and Communication
  Networks  \textbf{2020},  1–16 (Sep 2020). \doi{10.1155/2020/8858010}

\bibitem{Tokos2023}
Tokos, A.: Evaluating fuzzing tools for automated testing of rest apis using
  openapi specification  (2023), \url{https://www.doria.fi/handle/10024/187395}

\bibitem{Viglianisi2020}
Viglianisi, E., et~al.: Resttestgen: Automated black-box testing of restful
  apis. pp. 142--152. IEEE (10 2020). \doi{10.1109/ICST46399.2020.00024}

\bibitem{constrainedData}
Vitorino, J., et~al.: Constrained adversarial learning and its applicability to
  automated software testing: a systematic review (2023).
  \doi{10.48550/ARXIV.2303.07546}

\bibitem{Zheng2021}
Zheng, Y., et~al.: Automatic web testing using curiosity-driven reinforcement
  learning. p. 423–435. IEEE Press (2021). \doi{10.1109/ICSE43902.2021.00048}

\bibitem{Zhu2022}
Zhu, Y., et~al.: The application of neural network for software vulnerability
  detection: a review. Neural Computing and Applications  \textbf{35}(2),
  1279–1301 (Nov 2022). \doi{10.1007/s00521-022-08046-y}

\end{thebibliography}

\end{document}